\documentclass[twocolumn,amsmath,amssymb,prl,superscriptaddress,showpacs]{revtex4}
\usepackage{graphicx}
\usepackage{bm}

\begin{document}

\title{Graphene on incommensurate substrates:\\ trigonal warping and emerging Dirac cone replicas with halved group velocity}

\author{Carmine Ortix}
\affiliation{Institute for Theoretical Solid State Physics, IFW Dresden, D01171 Dresden, Germany}
\author{Liping Yang}
\affiliation{Institute for Theoretical Solid State Physics, IFW Dresden, D01171 Dresden, Germany}
\author{Jeroen van den Brink}
\affiliation{Institute for Theoretical Solid State Physics, IFW Dresden, D01171 Dresden, Germany}

\date{\today}

\begin{abstract}
The adhesion of graphene on slightly lattice-mismatched surfaces, for instance of hexagonal boron nitride (hBN) or Ir(111), gives rise to a complex landscape of sublattice symmetry-breaking potentials for the Dirac fermions.
Whereas a gap at the Dirac point opens for perfectly lattice-matched graphene on hBN, we show that the small lattice incommensurability prevents the opening of this gap and rather leads to a renormalized Dirac dispersion with a trigonal warping. This warping breaks the effective time reversal symmetry in a single valley. On top of this a new set of massless Dirac fermions is generated, which are characterized by a group velocity that is half the one of pristine graphene. 
 \end{abstract}

\pacs{73.22.Pr, 73.21.Cd, 72.80.Vp}
\maketitle

{\it Introduction} -- One of the main experimental challenges towards the realization of next-generation graphene electronics technology is the possibility to access the low energy Dirac point physics. Silicon oxide (SiO$_2$) substrates, for instance, are not ideal for graphene because of the trapped charges in the oxide. These impurity-induced charge traps  limit the device performances and make the low energy physics inaccessible \cite{mar07}. It has been recently shown that placing graphene on  hexagonal boron nitride (hBN) yields improved device performances \cite{dea10} -- graphene on hBN can have mobilities and charge inhomogeneities almost an order of magnitude better than graphene devices on SiO$_2$. 

hBN is interesting because it has the same honeycomb lattice structure of graphene, but only with two atoms in the unit cell, B and N, that are chemically inequivalent. Precisely this causes hBN to be a wide bandgap insulator.
When graphene is placed on top of a hBN surface, the lowest energy stacking configuration has one set of C atoms on top of B and the other C sublattice in the middle of the BN hexagons \cite{gio07,sac11}  -- assuming perfect lattice matching between graphene and hBN. 
Consequently the substrate-induced potential breaks the graphene sublattice symmetry. This leads to a gap at the Dirac point and hence a robust mass for the Dirac fermions. First principles band structure calculations \cite{gio07} put this gap at $\sim 50$ meV -- an energy roughly twice as large as $k_B T$ at room temperature. However, recent scanning tunneling microscopy experiments \cite{xue11,dec11} {\it do not}  detect a sizable bandgap. 

Within an effective continuum approach, here we show that this discrepancy originates from the 1.8 \% lattice mismatch \cite{liu03} between graphene and hBN  which leads to a  Moir\'e superstructure with periodicity much larger than the graphene lattice constant.  In this Moir\'e lattice, carbon atoms are embedded in a local environment of boron and nitrogen atoms that is varying continuously and periodically. This leads to a complex landscape of local sublattice symmetry-breaking terms which prevent the opening of a band gap at the Dirac point. Due to the incommensurability, the Dirac cones are instead preserved in renormalized form, with a threefold global symmetry due to a substrate-induced trigonal warping, which is in excellent agreement with experimental observations \cite{rus10}. In addition we also show that a new set of massless Dirac fermions is generated at the corners of the supercell Brillouin zone. These quasiparticles are characterized by a collinear group velocity $v_F$ which in the relevant weak coupling regime equals one half of $v_F^0$, the Fermi velocity in pristine graphene. As a set of these newly generated massless Dirac fermions does not overlap in energy with any other states, gated or doped graphene triangular Moir\'e superlattices provide a clear way to probe these 
Dirac fermions.  

Before presenting the calculations that explicate these results, we wish to point out that very similar physics arises for graphene on incommensurate substrates other than hBN, in particular for the experimentally relevant Moir\'e superlattices formed by graphene epitaxially grown on Ir(111) surfaces \cite{bal10,rus10}. As the (111) surface iridium atoms form a triangular lattice, there are two distinct local Ir-C configurations with a high symmetry \cite{vdb10}.  The first one occurs when a C atom is on top of Ir, situating its three neighbors in throughs between Ir sites -- the natural equivalence of the two triangular graphene sublattices is therefore broken. In the other high symmetry Ir-C configuration, the honeycomb carbon ring is centered above an iridium atom. In this case the effect of the Ir charges does not break the sublattice symmetry and therefore, no gap should open at the Dirac point.  While H decorated graphene/Ir(111) superstructures have been reported to give rise to absolute band gap openings \cite{bal10,vdb10}, recent angle-resolved photoemission spectroscopy data have rather shown an anisotropic behavior of the massless Dirac fermions close to the Dirac points due to an enhanced trigonal warping  \cite{rus10}. 

A number of interesting theoretical predictions exist on graphene superlattices. It is known that external one-dimensional periodic potentials can lead to a huge anisotropic renormalization of the electronic spectrum \cite{par08np,par08nl}, emerging zero modes \cite{bre09} and even to a  Landau-like level spectrum as a result of the presence of  extra Dirac points \cite{sun10}.  Dirac cone replicas at different ${\bf k}$ point in the Brillouin zone (BZ) have been reported also in triangular graphene superlattices \cite{par08} as well as in bilayer graphene superlattices \cite{kil11,tan11}. These findings, however, rely on a description that {\it disregards} local sublattice symmetry-breaking terms, which are crucial when investigating the opening/closing of gaps in graphene on slightly incommensurate hBN or Ir(111). In this Letter, we investigate the modification of the electronic spectrum of graphene Moir\'e  superlattices taking explicitly into account these essential, slowly varying, sublattice symmetry-breaking terms in an effective continuum approach. 

{\it Effective-Hamiltonian} -- We start out by taking into account the interaction induced by the substrate charges as an external electrostatic potential for graphene's Dirac electrons. The potential has a triangular periodicity that coincides with the arrangements of the centers of the BN hexagons (Ir atoms at the 111 surface): ${\it V}({\bf r})=\sum_{{\bf G}} V_{\bf G} \mathrm{e}^{i {\bf G} \cdot {\bf r}}$ where the ${\bf G}$'s are the reciprocal lattice vectors and $V_{\bf G}$ the corresponding amplitudes whose magnitudes depend on the modulus of ${\bf G}$ alone. 
In the following, we restrict the sum to the six wavevectors with equal magnitude $G=4 \pi / (3 \, a_S)$, ${\bf G} /G=\left(\pm 1, 0 \right)$, $\left(\pm \cos{\pi/3}, \pm \sin{\pi/3}\right)$ , where $\sqrt{3} \,a_S$ indicates the BN hexagons (Ir-Ir) distance. 
As the mismatch between $a_S$ and the graphene carbon-carbon distance $a$ is small, we can 
evaluate the effect of the substrate-induced electrostatic potential  on the two triangular graphene sublattices A/B  as the sum of products of rapidly varying parts $\exp{(i {\bf G}_{SR} \cdot {\bf r}_j^{A/B})}$,  $r_j^{A/B}$ being the actual atomic positions  and ${\bf G}_{SR}$  rescaled wavevectors with magnitude $G_{SR}=4 \pi / (3 \, a)$, times  slowly varying parts $\exp{(i \widetilde{{\bf G}} \cdot {\bf r})}$ which we treat in the continuum limit. The $\widetilde{{\bf G}}$'s  are the rescaled "coarse-grained" wavevectors with magnitude $\widetilde{G}=4 \pi |\delta a| / 3 a^2$ where $\delta a= a_S - a$ indicates the lattice mismatch which without loss of generality has been assumed positive.
As a result, the effect of the substrate charges leads to an average external potential acting equally on each carbon atom and a mass term breaking the graphene sublattice symmetry given by
${\cal {\it V}}_{\pm} ({\bf r})=[{\it V}_{A} ({\bf r}) \pm {\it V}_{B} ({\bf r})] / 2 = V_0 / 2 \sum_{\widetilde{{\bf G}}}  \left[1\pm \exp{i \phi_{\widetilde{\bf G}}} \right] \times \exp{(i \widetilde{\bf G} \cdot {\bf r} )}$. The contribution of the rapidly varying parts of the potential are now encoded in the non-trivial phase factors $ \phi_{\widetilde{\bf G}}=\widetilde{\bf G} / \widetilde{G} \cdot {\boldsymbol \delta} / a$ where $\boldsymbol{\delta}= - a (1,0)$ is the graphene nearest-neighbor vector. 

Since the large periodicity of the Moir\'e superstructure prevents intervalley scattering, 
we can describe the low-energy quasiparticles 
near the corners ${\bf K}_{\pm}=\left\{2 \pi / (3 a), \pm 2 \pi / (3 \sqrt{3} a)\right\}$ of the graphene hexagonal Brillouin zone 
as 4-dimensional spinors $\Psi=\left[\psi_{{\bf K}_{+}, A}, \psi_{{\bf K}_{+}, B},\psi_{{\bf K}_{-}, B},\psi_{{\bf K}_{-}, A}\right]$, characterizing the electronic amplitudes on the two crystalline sublattices,  with an effective Hamiltonian 
\begin{equation}
\hat{{\cal H}}= v_F^0 \tau_0 \otimes  {\bf p} \cdot {\boldsymbol \sigma} + {\it V}_{+}({\bf r}) \tau_0 \otimes \sigma_0 + {\it V}_{-}({\bf r}) \tau_z \otimes \sigma_z.
\label{eq:hamiltonian}
\end{equation}
Here we use direct products of Pauli matrices $\sigma_{x,y,z}$,  
$\sigma_0 \equiv \hat{1}$ 
acting in the sublattice space and $\tau_{x,y,z}$,$\tau_0 \equiv \hat{1}$ acting on the valley degree of freedom (${\bf K}_{\pm}$). 
 \begin{figure}
\includegraphics[width=\columnwidth]{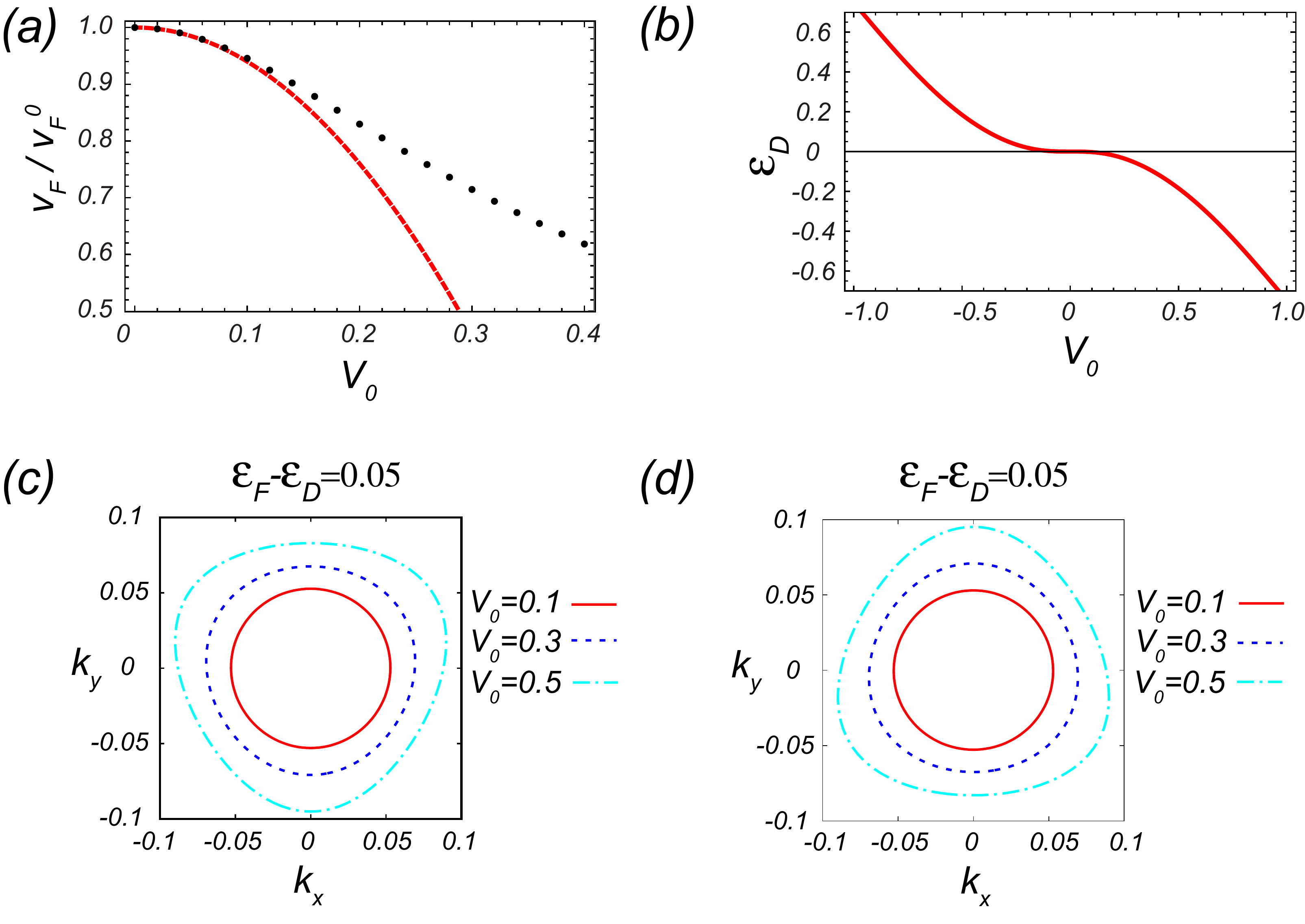}
\caption{(Color online) 
(a) Renormalization of the Fermi velocity at the Dirac point as a function of $V_0$.  The points are the results of the exact diagonalization of the low-energy Hamiltonian whereas the continuous line is the analytical result from second-order perturbation theory.
(b) Energy of the Dirac point as a function of the external potential amplitude $V_0$.
(c),(d) Fermi lines  for different values of the potential amplitude $V_0$ close to the ${\bf K}_{+}$ 
(c) and the ${\bf K}_{-}$(d) valleys. Energies and wave vectors are in units of $\hbar v_F \widetilde{G}$ and $\widetilde{G}$ respectively. 
}
\label{fig:fig2}
\end{figure}
For ${\it V}_{\pm}({\bf r}) \equiv 0$, $\hat{\cal H}$ has a chiral symmetry which can be expressed as $\left\{\hat{\cal H}, \tau_0 \otimes \sigma_z \right\} \equiv 0$. This anticommutation relation implies that in each valley  any eigenstate $\Psi_{\epsilon}$ with energy $\epsilon$ has a particle-hole partner  $ \tau_0 \otimes \sigma_z \Psi$ with energy $-\epsilon$. 
This property implies  the doubly degeneracy of the zero energy states in each valley. 
In the presence of substrate-induced interactions of the form as in Eq.~\ref{eq:hamiltonian}, 
the system still possesses a chiral symmetry provided the external superlattice potentials satisfy ${\it V}_{\pm}({\bf r}+{\bf T})=-{\it V}_{\pm}({\bf r})$. In this case it is possible to define a new chiral operator \cite{sun10} $ \tau_0 \otimes \sigma_z {\cal S}$ where ${\cal S}$ is a shift operator ${\cal S} \Psi ({\bf r}) = \Psi ({\bf r}+{\bf T})$.  For the electrostatic potentials defined above, 
a translation vector for which 
the triangular  potential 
${\it V}_{+}({\bf r}+{\bf T}) \equiv -{\it V}_{+}({\bf r})$ 
is absent 
thereby implying  particle-hole symmetry breaking and a consequent lifting  of the zero energy states degeneracy. 
\begin{center}
\begin{figure*}
\includegraphics[width=\textwidth]{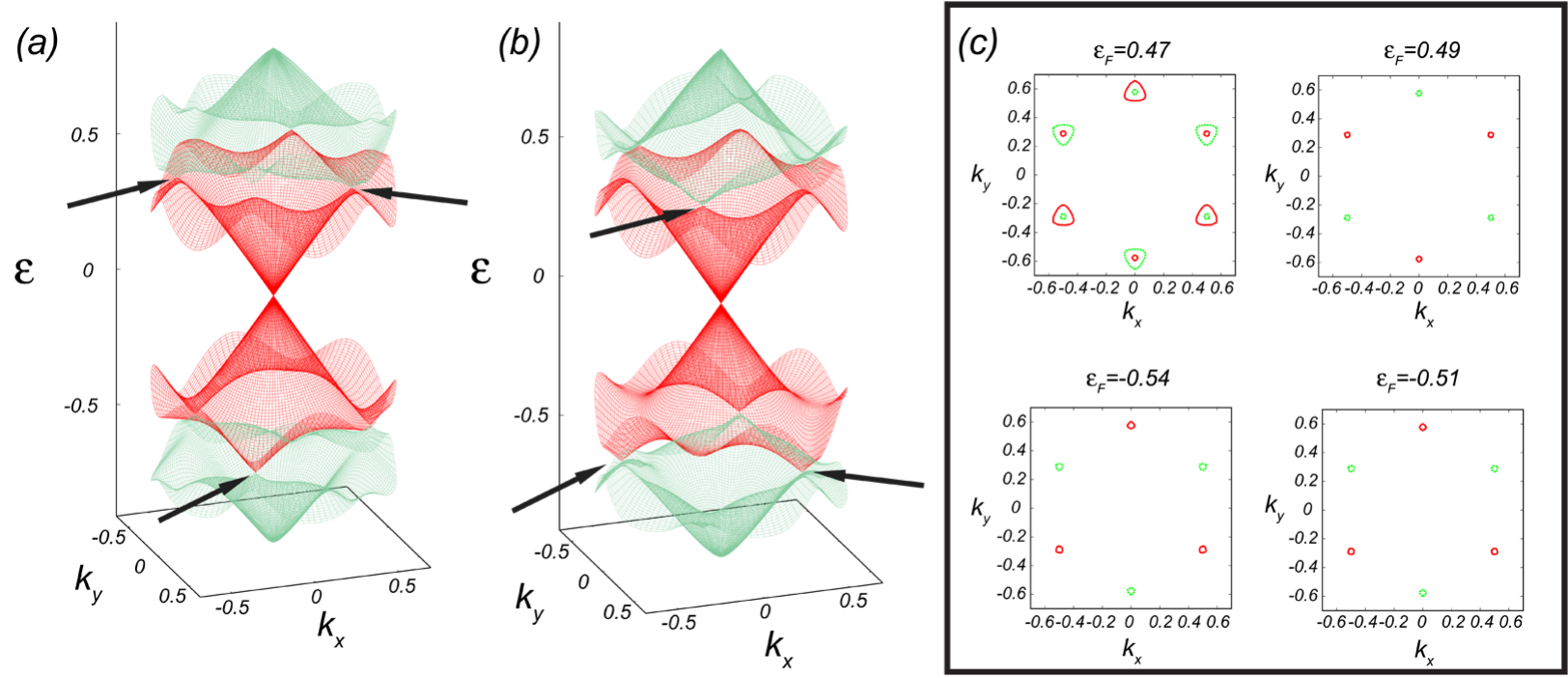}
\caption{ 
(Color online) Energy dispersion relations for a graphene triangual Moir\'e superlattice with $V_0=0.1 \hbar \,v_F^0 \, \widetilde{G}$ close to the two graphene valleys ${\bf K}_{+}$ (a) and ${\bf K}_{-}$ (b). The arrows indicate the corners $\widetilde{{\bf K}}_{\pm}$ of the SBZ where the new Dirac cones are generated . (c) Topology of the Fermi lines close to the emergent Dirac cones. Units are the same as in Fig.\ref{fig:fig2}.}
\label{fig:fig1}
\end{figure*}
\end{center}
This, however, does not lead to the opening of any absolute band gap since the $\widetilde{{\bf G}} \equiv 0$ component of the local sublattice symmetry breaking terms identically vanishes. Therefore, the Dirac cones are preserved with the effect of particle-hole asymmetry eventually leading to a shift of the conical points [shown in Fig.\ref{fig:fig2}(b)] of the two valleys reminiscent of the graphene doping caused by adsorption  of metal substrates \cite{gio08}. 
We also find the Dirac cones to be renormalized in  triangular Moir\'e superlattices. In Fig.\ref{fig:fig2}(a) we show the behavior of the collinear Fermi velocity at the conical points for different values of the interaction strength $V_0$. The substrate-induced interaction leads to a decrease of the Fermi velocity as can be found in the weak potential limit by treating the effect of the electrostatic potentials in second-order perturbation theory [c.f. continuous line in Fig.\ref{fig:fig2}(a)] according to which  
$$v_F=v_F^0 \left[1- \dfrac{6 V_0^2}{\hbar^2 v_F^{0\,2} \widetilde{G}^2} \right].$$ 

The local sublattice symmetry-breaking term breaks the effective time-reversal symmetry on a single valley  \cite{bee08},
$\widetilde{{\cal T}}=i \left(\tau_0 \otimes \sigma_y \right) {\hat {\cal C}}$, 
with ${\hat {\cal C}}$ the operator of complex conjugation and 
$\widetilde{{\cal T}}\left[\psi_{{\bf K}_{+}, A}, \psi_{{\bf K}_{+}, B},\psi_{{\bf K}_{-}, B},\psi_{{\bf K}_{-}, A}\right]=[\psi^{\star}_{{\bf K}_{+}, B}, -\psi^{\star}_{{\bf K}_{+}, A},\psi^{\star}_{{\bf K}_{-}, A},-\psi^{\star}_{{\bf K}_{-}, B}]$. 
This is clearly visible in Fig.\ref{fig:fig2}(c),(d) 
where we show the topology of the Fermi lines close to the Dirac points in the two graphene valleys. There is  a trigonal warping 
which breaks the ${\bf k} \rightarrow -{\bf k}$ symmetry of the Fermi lines, {\it i.e.}  $\epsilon({\bf K}_{\pm}, {\bf k}) \neq \epsilon({\bf K}_{\pm}, -{\bf k})$, 
consistent with the threefold symmetry of the bandstructure  experimentally detected in Ir(111) superlattices \cite{rus10}. 
The trigonal warping has an opposite effect on the two valleys since the external electrostatic potentials {\it do not} break  the true time-reversal symmetry interchanging the valleys \cite{bee08} ${\cal T}=-\left(\tau_y \otimes \sigma_y \right) {\hat {\cal C}}$ with ${\cal T} \left[\psi_{{\bf K}_{+}, A}, \psi_{{\bf K}_{+}, B},\psi_{{\bf K}_{-}, B},\psi_{{\bf K}_{-}, A}\right] = [\psi^{\star}_{{\bf K}_{-}, A}, -\psi^{\star}_{{\bf K}_{-}, B},-\psi^{\star}_{{\bf K}_{+}, B},\psi^{\star}_{{\bf K}_{+}, A}]$.  Hence, the Fermi lines fulfill $\epsilon({\bf K}_{\pm}, {\bf k}) \equiv\epsilon({\bf K}_{\mp}, -{\bf k})$ as can been shown in the weak potential limit where the energy dispersion of the low-energy quasiparticles reads 
\begin{equation}
\epsilon({\bf K}_{\pm}, {\bf k}) \simeq   \hbar v_F s |k| -  \dfrac{3 s}{\hbar v_F^0} \dfrac{|k|^2}{\widetilde{G}^3} V_0^2 \left[ 1 \pm  \sqrt{3} \sin{3 \theta_k}\right].
\end{equation}
Here $\theta_k$ is the angle of the vector ${\bf k}$ with respect to the ${\hat k}_x$ direction and $s=\pm 1$ for quasi-electrons and quasi-holes respectively. By increasing the strength of the potential $V_0$ [c.f. Fig.\ref{fig:fig2}(c),(d)], the warping effect is enhanced and results in an anisotropy much larger than the one  expected in freestanding graphene \cite{rus10}.

{\it Emergence of Dirac cone replicas} -- We have also obtained the energy dispersion in the full supercell Brillouin zone (SBZ) by exact diagonalization of the Hamiltonian Eq.~\ref{eq:hamiltonian}. The effect of the substrate-induced external potential ${\it V}_{\pm}({\bf r})$ has been incorporated into our calculations through the scattering matrix elements between the chiral eigenstates of the graphene quasiparticles 
\begin{equation*}
\psi_s({\bf K}_{\pm}, {\bf k})= \left( \begin{array}{c} \psi_{{\bf K}_{\pm}, A} \\ \psi_{{\bf K}_{\pm}, B} \end{array} \right)= \dfrac{1}{\sqrt{2}} \left( \begin{array}{c} 1 \\  s \,\mathrm{e}^{\pm i \, \theta_{k}} \end{array} \right) \mathrm{e}^{i \, {\bf  k} \cdot {\bf r}}.
\end{equation*}
Fig.\ref{fig:fig1} shows the ensuing energy dispersion of the first and second bands above and below the original Dirac points in each valley. Contrary to Ref.~\onlinecite{par08} we do not find a generation of Dirac cones at the six $\widetilde{{\bf M}}$ points of the SBZ. Indeed, the energy separation between the first and the second band above and below the original Dirac points goes to zero respectively at the $({\bf K}_{\pm},\widetilde{{\bf K}}_{\pm})$,  $({\bf K}_{\pm},\widetilde{{\bf K}}_{\mp})$ corners of the SBZ. 
This qualitative difference is caused by the sublattice symmetry-breaking term ${\it V}_{-}({\bf r})$ which  
does not allow for a sixfold symmetry of the bandstructure. 
The topology of the Fermi lines close to the SBZ corners clearly shows the emergence of Dirac cone replicas. 
However, while above the original Dirac points [c.f. Fig.\ref{fig:fig1}(c)] 
these new massless quasiparticles are obscured by other states, below the original Dirac points there is an energy window where there are no other states than the new massless Dirac fermions. 
Therefore there is one energy value -- apart from the original Dirac point -- where the density of states (DOS) vanishes linearly. 
It is worth noticing that this asymmetry of the DOS reflects the particle-hole symmetry breaking discussed above. 

\begin{figure}
\includegraphics[width=\columnwidth]{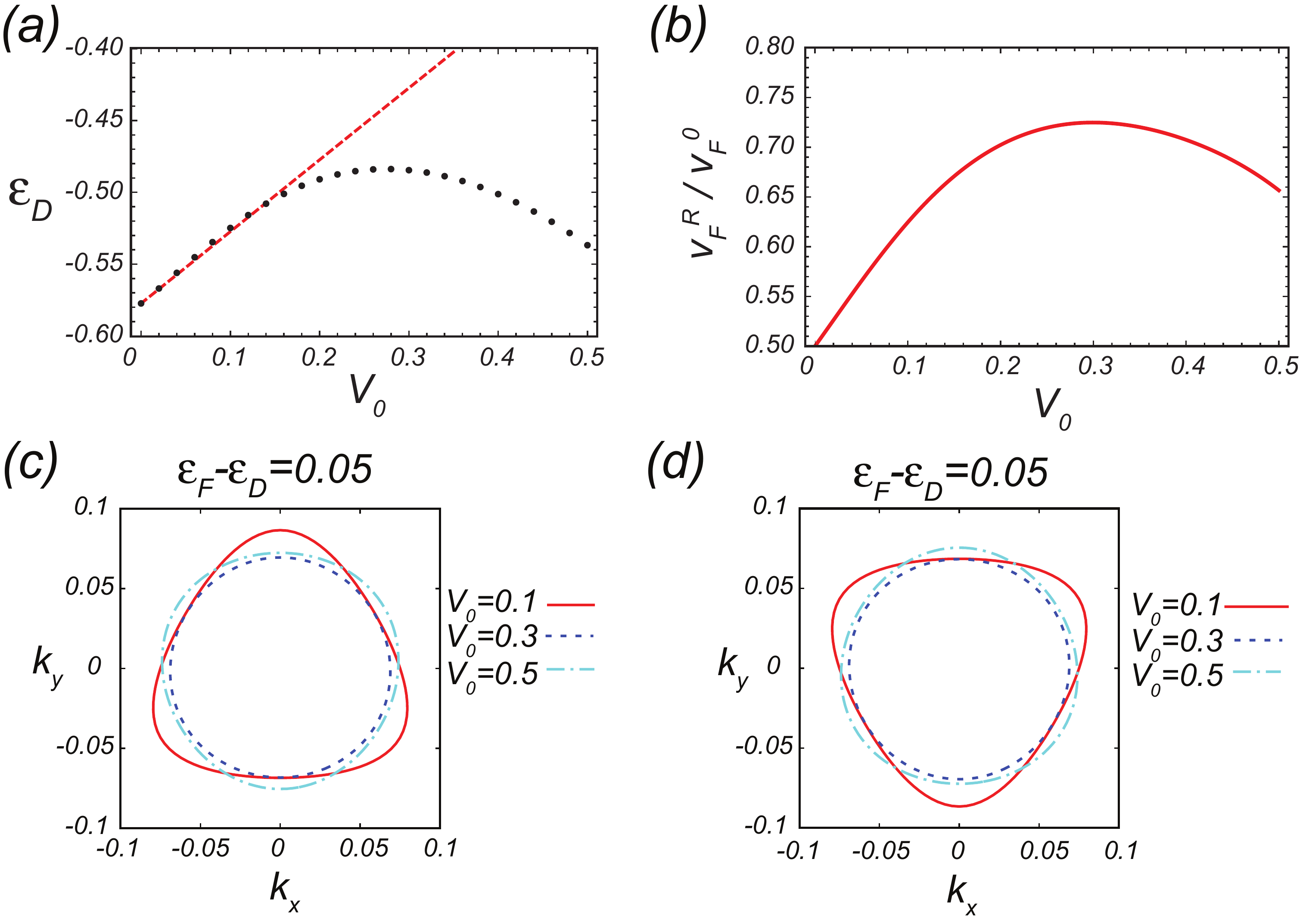}
\caption{ (Color online) Behavior of the Dirac point energy  $\epsilon_{D}({\bf K}_{\pm}, \widetilde{{\bf K}}_{\mp})$ (a)  as a function of the external potential amplitude $V_0$. The continuous line is the result of the degenerate perturbation theory whereas the points are the result of the exact diagonalization. (b) Group velocity at the Dirac point replicas as function of the strength of the potential $V_0$. (c),(d)  Fermi lines close to the emergent Dirac points
 $ \epsilon_{D}({\bf K}_{+}, \widetilde{{\bf K}}_{-})$  (c) and  $ \epsilon_{D}({\bf K}_{-}, \widetilde{{\bf K}}_{+})$ for different values of the amplitude $V_0$.
}
\label{fig:fig3}
\end{figure}

Further insight into the properties of the Dirac cone replicas is obtained by introducing an effective Hamiltonian close to the three equivalent corners of the SBZ \cite{gui10}. In the following we will restrict to consider the behavior close to the $({\bf K}_{\pm}, \widetilde{{\bf K}}_{\mp})$ points, relevant for the Dirac cone replicas generated below the original Dirac points. 
In the absence of external electrostatic potentials ${\it V}_{\pm}({\bf r}) \equiv 0$, there are three degenerate hole excitations with energy $\epsilon({\bf K}_{\pm}, \widetilde{{\bf K}}_{\mp})= -\hbar v_F \widetilde{K}$. 
This degeneracy is lifted by the substrate-induced electrostatic potentials and, as a result, one finds a singlet excitation with energy $\epsilon_{S}({\bf K}_{\pm}, \widetilde{{\bf K}}_{\mp})= -\hbar v_F \widetilde{K} - V_0$ and a doubly degenerate state with energy $\epsilon_{D}({\bf K}_{\pm}, \widetilde{{\bf K}}_{\mp})= -\hbar v_F \widetilde{K} + V_0 / 2 $. 
It can be easily shown that the effective Hamiltonian in the vicinity of this doubly degenerate state corresponds precisely to a massless two-dimensional Dirac equation with Fermi velocity $v_F^R=v_F^0 / 2$ and an isotropic dispersion relation 
\begin{equation}
\epsilon_{D}({\bf K}_{\pm}, {\boldsymbol \delta} {\bf k}_{\pm})= \epsilon_{D}({\bf K}_{\pm}, \widetilde{{\bf K}}_{\mp}) + \hbar \dfrac{v_F^0}{2}  s^{\prime}  \delta k_{\pm}
\end{equation}
where we introduced  ${\boldsymbol \delta} {\bf k}_{\pm}  = {\bf k}-\widetilde{{\bf K}}_{\mp}$ 
and $s^{\prime}=\pm 1$ is the band index. 
The foregoing weak coupling analysis is in perfect agreement with the numerical results obtained by exact diagonalization of the Hamiltonian. This is shown in Fig.\ref{fig:fig3}(a),(b) where we plot the behavior of the Dirac point energy $\epsilon_{D}({\bf K}_{\pm}, \widetilde{{\bf K}}_{\mp})$ and the Fermi velocity $v_F^R$ as a function of the potential strength $V_0$. 
Away from but close to the Dirac points, we find a trigonal warping  respecting the threefold symmetry of the band structure as it is  shown in Fig.\ref{fig:fig3}(c),(d) 
where we plot the Fermi lines for different values of $V_0$. 

{\it Conclusions} -- By setting up an effective continuum approach, we have derived the electronic properties of graphene Moir\'e superlattices generated by adhesion of graphene sheets onto lattice mismatched substrates. While the complex landscape of sublattice symmetry-breaking terms prevents the opening of a bandgap at the Dirac point, we have demonstrated that with mismatched substrates one can tailor the low-energy band dispersion. In agreement with recent experiments \cite{rus10}, we have found a threefold-symmetry of the band structure associated to a substrate-induced trigonal warping of the Dirac cones. In addition a new set of Dirac fermions is generated in graphene Moir\'e superlattices. By properly and explicitly accounting for local sub-lattice symmetry terms, we have shown that these new quasiparticles are generated at the corners of the supercell Brillouin zone and are characterized by a collinear group velocity at the conical points of $\sim v_F^0 / 2$. 

L.Y. thanks F. Guinea  for valuable discussions. This research was supported by the Dutch Science Foundation NWO/FOM.

\end{document}